\documentclass[prl,twocolumn,superscriptaddress,amsmath,showpacs]{revtex4} 

\usepackage{graphicx}
\usepackage{dcolumn}
\usepackage{bm}

\begin{document}


\title{Mesoscopic Speckle}

\author{Sheng Zhang}
\affiliation{Department of Physics, Queens College, The City University of New York, Flushing, NY 11365, USA}
\affiliation{ChiralPhotonics Inc., Pine Brook, NJ 07058, USA}
\author{Yitzchak Lockerman}%
\affiliation{Department of Physics, Queens College, The City University of New York, Flushing, NY 11365, USA}%
\author{Azriel Z. Genack}%
\affiliation{Department of Physics, Queens College, The City University of New York, Flushing, NY 11365, USA}%

\date{\today}

\begin{abstract}
We have measured the local and global statistics of singularity velocity, $v$, and have related these through the spatial correlation function of $v$. The distribution of $v$ is a mixture of a mesoscopic distribution of global change in the speckle pattern and the distribution for $v$ for Gaussian random fields. When $v$ is normalized by the standard deviation of the fractional intensity change, probability distributions and correlation function of $v$ approach those for random Gaussian fields. These results are directly analogous to the statistics of transmitted intensity normalized by the total transmission and provide a unified framework for understanding statistics of speckle evolution and intensity.
\end{abstract}

\pacs{42.25.Dd, 42.25.Bs, 42.30.Ms }

\maketitle

Coherent waves multiply scattered by a disordered medium form a speckled intensity pattern built upon a network of phase singularities at nulls of intensity. Current flux vortices circulate about these phase singularities towards which equiphase lines converge. \cite{1,2} The motion of these singularities is a sensitive probe of changes in the scattering material and can be used to detect material defects and deformation, \cite{3} to diagnose cardiac fibrillation, \cite{4} and to precisely position  small particles. \cite{5} Field structures within speckle patterns are generic in that these arise in a wide variety of circumstances, \cite{A} are robust under perturbation and are governed by universal statistics. \cite{2,Freund,6} The scale of the speckle pattern depends only upon the angular distribution of the scattered field. Perhaps because of the universality of the speckle pattern, its structural statistics have only recently been utilized to investigate the character of the wave within the medium. Studies of mesoscopic fluctuations have focused on first and second order intensity statistics in space, \cite{B} frequency, \cite{C} and time, \cite{11a} and on their relation to enhanced fluctuations of total transmission \cite{E} and conductance \cite{F} and to Anderson localization. \cite{G} However, recent studies have found that mesoscopic fluctuations and localization strongly influence the statistics of the structure of static \cite{6} and evolving \cite{6d} speckle patterns. The statistics of various measures of overall change of the transmitted speckle pattern as the frequency is tuned were found to be governed by a distribution with the same functional form as that for total transmission and to similarly depend only upon the variance of the distribution itself. Among the measures of speckle change are the average displacement or speed of singularities, and the standard deviations of fractional intensity change or phase change within the speckle pattern. The similarity in these statistics is surprising since they are oppositely correlated with resonances with quasimodes of the medium. Whereas the average intensity within a speckle pattern for a single incident mode $a$, $I_a$, which is the total transmission, peaks on resonance, the speckle change is maximized between resonances. On resonance, the speckle pattern is essentially that of the resonant mode and so changes slowly with frequency shift.

In this Letter, we report striking parallels between the spatial correlation of local speckle pattern change, measured by the generalized speed of phase singularities at the sample output with frequency shift, $v \equiv dr_s/d\nu$, and intensity, $I$. We measure the probability distribution, $P(\tilde{v})$, of $v$ normalized by its ensemble average, $\tilde{v} = v/\langle v\rangle$, and the spatial cumulant correlation function, $C_v(\Delta r) = \langle \delta \tilde{v}({\bf r}) \tilde{v}({\bf r}+\Delta {\bf r}) \rangle$, where $\delta \tilde{v} = \tilde{v}-1$, $\Delta r$ is the distance between the two singularitieslocated at ${\bf r}$ and ${\bf r}+\Delta {\bf r}$. $P(\tilde{v})$ broadens and the long-range component of $C_v(\Delta r)$, $\kappa_v$, increases in the localization transition. $\kappa_v$ is nearly equal to the variance of a key measure of change of the speckle pattern as a whole, which is the standard deviation of fractional intensity change within the speckle pattern. This is similar to the near equality of the degree of intensity correlation $\kappa_I$ to the variance of total transmission. When $v$ is normalized by $\eta$, a quantity which characterizes the whole speckle change, and $I$ is normalized by $I_a$, their corresponding statistics closely match those for Gaussian random wave fields.

We measured the microwave field transmitted through samples of alumina spheres contained in a 61-cm-long copper tube with the diameter of 7.0 cm. The sample is composed of 0.95-cm-diameter alumina spheres with refractive index 3.14 embedded in Styrofoam shells of refractive index 1.04 to produce an alumina volume fraction of 0.068. \cite{11} The in- and out-of-phase components of the transmitted field polarized along a 4-mm-long and 0.5-mm-diameter wire antenna are measured with use of a vector network analyzer. The spatial distribution of the transmitted field over a range of frequencies is obtained by measuring field spectra at each point on a 1-mm-square grid over the output surface of the sample. Measurements are made over the frequency ranges 14.7-15.7 GHz and 10-10.24 GHz in which waves are diffusive and localized, respectively. Frequency steps are chosen to be approximately 1/7 of the field correlation frequency. Measurements were made in 40 and 71 different configurations for diffusive and localized waves, respectively. 

In order to accurately determine the positions of phase singularities, the 2D sampling theorem is applied to the data to reconstruct the speckle patterns on a $50\mu m \times 50\mu m$ grid. \cite{6} The sampling theorem is also used to interpolate in the frequency domain, so that spectra with 120kHz and 250kHz frequency steps are obtained for localized and diffusive waves, respectively. This allows us to accurately locate phase singularities and to measure the magnitude $v$. Velocity distributions for diffusive and localized waves are shown in Fig. 1(a). These results are compared to simulations for Gaussian random waves generated by the superposition of 300 phased plane waves, $E(x,y,z)=\sum_i A_i \exp [i(k_xx+k_yy+k_zz)]$. Each of the components of the $k$-vector and the amplitude $A_i$ are drawn from a Gaussian distribution. We noticed, however, the simulation result is not in good agreement with the theoretical formula, $P(\tilde{v}) = \frac{8\pi^2\tilde{v}}{(\pi^2\tilde{v}^2+4)^2}$, derived from Gaussian random waves in \cite{2}. The singularity velocity in the $x-y$ plane is tracked as $z$ increases. For diffusive waves, $P(\tilde{v})$ is close to the results of the simulations, while $P(\tilde{v})$ for localized wave is noticably broader. Measurements of $C_v(\Delta r)$ for diffusive and localized waves are shown in Fig. 1(b). No corresponding theoretical calculation has been done so far. Extremely high value of $C_v(0)$ is consistent with the fact that the velocity of singularities diverges as $\Delta r \rightarrow 0$ where singularities are created or annihilated. \cite{2} $C_v(\Delta r)$ falls rapidly with $\Delta r$ and reaches a constant value denote by $\kappa_v$, which is 0.039 for diffusive and 0.648 for localized waves, respectively. Thus the correlation function can be expressed as the sum of a short-range term and a constant: $C_v(\Delta r) = C_{v,short}(\Delta r) + \kappa_v$.

\begin{figure}
\includegraphics[scale=0.7]{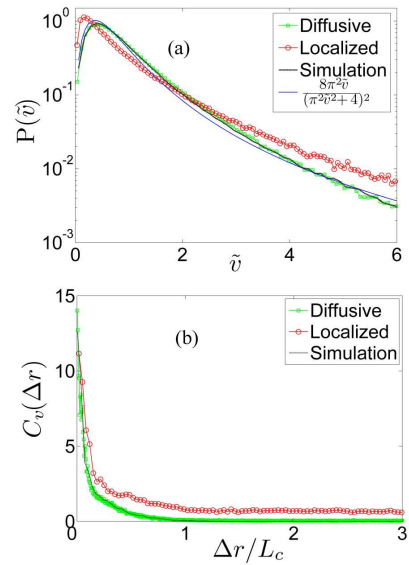}
\caption{\label{Fig1} (Color Online) (a) The probability distributions and (b) cumulant spatial correlation functions of velocity of phase singularities normalized to the respective ensemble averages, $\tilde{v}$, for diffusive and localized waves.}
\end{figure}

The statistics of singularity velocity can be compared to first and second order statistics of polarized intensity which are plotted in Fig. 2(a) and (b). Fluctuations of I are seen greatly enhanced for localized waves. The structure of $C_v(\Delta r)$ and of the cumulant correlation function for intensity, $C_I(\Delta r)$, which is plotted in Fig. 2(b) using the same data used in Fig. 1 are similar in that $C_I(\Delta r)$ may be expressed as, $C_I(\Delta r) = C_{I,short}(\Delta r) + \kappa_I$. Here $\kappa_I$ is the degree of correlation, which is the value of $C_I(\Delta r)$ at points at which the field correlation functions vanishes. \cite{PRL2002} $\kappa_I = 0.12$ for the diffusive and 3.0 for localized waves. In quasi-1D samples with a large number of transverse modes, the field in individual speckle patterns can be assumed to be a Gaussian random variable. Thus the probability distribution of intensity normalized by the average intensity within the speckle pattern, $I^\prime(r) = I(r)/I_a$, should be statistically independent of the total transmission and the polarized intensity should follow the probability distribution $P(\tilde{I}^\prime)=\exp(-\tilde{I}^\prime)$. The measured $P(\tilde{I}^\prime)$, however, deviates from this prediction [see Fig. 2(c)]. Such deviations are expected because the number of transverse waveguide modes is small; approximately 30 at 10GHz and 50 at 15 GHz. Agreement with Gaussian statistics is better for diffusive waves since the number of modes is larger. If we assume $\tilde{I}^\prime$ and $I_a$ are statistically indepentent, we find $\langle I^\prime(r)I^\prime(r+\Delta r)I^2_a \rangle = \langle I^\prime(r)I^\prime(r+\Delta r)\rangle \langle I^2_a\rangle $. The cumulant intensity correlation function, $C_{I^\prime}=\Gamma_I(\Delta r)-\langle I \rangle^2$, can then be expressed as,
\begin{equation}\label{eq1}
  C_I(\Delta r) = \left[ C_{I^\prime}(\Delta r) +1 \right] \left[ {\rm var}(\tilde{I}_a) +1 \right] -1
\end{equation}
$C_{I^\prime}(\Delta r)$ and the square of the corresponding field correlation function, $F(\Delta r)$, are calculated for diffusive and localized waves and seen to be similar in Fig. 2(d). Since $C_{I^\prime}(\Delta r) \rightarrow 0$ for large $\Delta r$, Eq.(\ref{eq1}) gives, $\kappa_I = {\rm var}(\tilde{I}_a)$. This is roughly consistent with the measured values of ${\rm var}(\tilde{I}_a)$ of 0.14 for diffusive and 3.3 for localized waves (see Table 1). Again, $C_{I^\prime}(\Delta r)$ is closer to $F(\Delta r)$ for diffusive waves rather than for localized waves showing the corresponding field statistics is closer to Gaussian when the number of transverse modes is larger.

\begin{figure}
\includegraphics[scale=0.7]{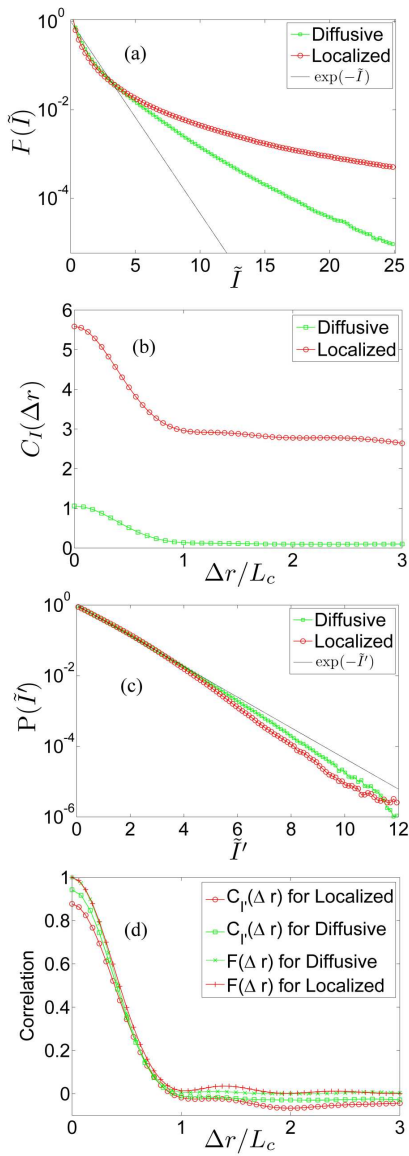}
\caption{\label{Fig2} (Color Online) The first and second order statistics of $\tilde{I}$ and $\tilde{I}^\prime$ for diffusive and localized waves. (a) Probability distributions of $\tilde{I}$. (b) Cumulant correlation functions of $\tilde{I}$. The saparations $\Delta r$ are normalized by the corresponding correlation lengths $L_c$, which are the first zeros of the real part of the field correlation functions. (c) Probability distributions of $\tilde{I}^\prime$. (d) Cumulant correlation functions of $\tilde{I}^\prime$ and their comparison to the square of the field correlation functions.}
\end{figure}

The character of speckle pattern change can also be traced to the combined factors of long-range correlation in speckle change and the statistical independence of local and global fluctuations. We consider the  motion of phase singularities and assume that the global change of speckle patterns, denoted by $\eta$, is statistically independent of the velocity of individual singularities normalized by this change, $v^\prime = v/\eta$. $C_v(\Delta r)$ can then be expressed as,
\begin{equation}\label{eq2}
  C_v(\Delta r) = \left[ C_{v^\prime}(\Delta r) +1 \right] \left[ {\rm var}(\tilde{\eta}) +1 \right] -1
\end{equation}
If the local changes in the speckle patterns were a Gaussian random process, we would expect that $C_{v^\prime}(\Delta r) \rightarrow 0$ for large $\Delta r$, since the fields in distant regions would not be correlated. In the limit of large $\Delta r$, this would give $\kappa_v = {\rm var}(\tilde{\eta})$.

Several candidates for a paremeter $\eta$ to quantify the change of the speckle pattern were discussed in \cite{6d}, including the average velocity of phase singularities $v_a$, the standard deviation of phase changes $\sigma_{\Delta\varphi}$ and the standard deviation of fractional intensity change, $\sigma_{\Delta I^*}$, where $I^* = \frac{I(\nu+\Delta\nu) - I(\nu)}{I(\nu+\Delta\nu) + I(\nu)}$. The spectra of these quantities in specific sample configurations are similar for localized waves and their probability distributions have the same functional form as the probability distribution of total transmission, though the variances of the distributions differ. \cite{6d} The measured variances of $v_a$, $\sigma_{\Delta\varphi}$ and $\sigma_{\Delta I^*}$ together with $\kappa_v$ for both diffusive and localized waves are given in Table 1. We find that ${\rm var} (\tilde{\sigma}_{\Delta I^*})$ is closest to $\kappa_v$ in both cases, while ${\rm var} (\tilde{v}_a)$ and ${\rm var} (\tilde{\sigma}_{\Delta\varphi})$  are higher. This is also reflected from considerably smoother spectra of $\tilde{\sigma}_{\Delta I^*}$ than those for $\tilde{v}_a$ and $\tilde{\sigma}_{\Delta\varphi}$, which have been shown in \cite{6d}. In the limit $\Delta\nu \rightarrow 0$, $\Delta I^*/\Delta\nu$ does not diverge near singularities as does $\Delta\varphi/\Delta\nu$ \cite{2}. Thus, ${\rm var} (\sigma_{\Delta I^*})$ more reliably reflects the change of the speckle pattern as a whole, while $\sigma_{\Delta\varphi}$ is strongely effected by the immediate region around the singularity and $v_a$ is a property of the small numbers of singularities. This suggests that $\eta = \sigma_{\Delta I^*}$ is a more practical choice as an indicator of global speckle change. 

\begin{table}
\centering
\begin{tabular}{c c c c c c c c}
\hline
  & $\kappa_I$ & ${\rm var}(\tilde{I}_a)$ & $\kappa_I$ & ${\rm var}(\tilde{v}_a)$ &${\rm var}(\tilde{\sigma}_{\Delta\varphi})$ &${\rm var}(\tilde{\sigma}_{\Delta I^*})$ & ${\rm var} (\tilde{v}^\prime_a)$ \\
\hline
 Diff & 0.12 & 0.14 & 0.039 & 0.193 & 0.087 & 0.045 & 0.131 \\
 Loc & 3.0 & 3.3 & 0.648 & 1.240 & 0.743 & 0.586 & 0.267  \\
\hline
\end{tabular}
\caption{Comparison of $\kappa_I$ and $\kappa_v$ to variance of global measures of intensity and speckle changes.}
\end{table}

When $v$ is normalized by $\eta = \sigma_{\Delta I^*}$ first and second order velocity statistics for diffusive and localized waves collapse to the results found in simulations for Gasussian waves (Fig. 3), just as was found for statistics of the intensity normalized by $I_a$, $I^\prime$. Despite many similarities, there are key differences in the relationships between global and local statistics for $v$ and $I$. Unlike intensity, which is defined at all points, there are only a small number of singularities in the speckle pattern; 15 for diffusive and 10 for localized waves on average. This difference leads to another dissimilarity between the statistics of $v$ and $I$, that ${\rm var}(\tilde{I}_a)$ is slightly larger than $\kappa_I$, whereas, ${\rm var}(\tilde{v}_a)$ is significantly greater than $\kappa_v$ (see Table 1). The source of this difference can be seen by first considering the relationship of ${\rm var}(\tilde{I}_a)$ and $\kappa_I$. The variance of total transmission can be expressed in terms of the spatial correlation function of intensity,
\begin{eqnarray}
  {\rm var}(\tilde{I}_a) &=& \frac{1}{A} \int_A C_I(\Delta r) d{\Delta r}^2  \nonumber\\
    &=& \frac{1}{A} \int_{A,short} C_{I,short}(\Delta r) d{\Delta r}^2 + \kappa_I  \label{eq3}\\
    &=& \Gamma_{I,short} + \kappa_I, \nonumber
\end{eqnarray}
where, $A$ is the total area of the output surface. The small excess of ${\rm var}(\tilde{I}_a)$ over $\kappa_I$, indicates that the assumptions made in Eq. (1) are not strictly valid. A quantitative measure of the breakdown of independence of $I\prime$ and $I_a$ is the relative magnitudes of the contributions to ${\rm var}(\tilde{I}_a)$ by the integral of the short-range correlation function $C_{I,short}(\Delta r)$ over the output surface and $k_I$. Since $C_{I,short}(\Delta r)$ falls rapidly to 0 for $\Delta r>L_C$ and the correlation length $L_C$ is much smaller than the diameter of the sample cross-section, the integral over $A$, giving $\Gamma_{I,short}=0.015$ for diffusive and 0.088 for localized waves, is significantly smaller than the corresponding values of $\kappa_I$. Thus the assumptions made are approximately valid and ${\rm var}(\tilde{I}_a)\approx \kappa_I$, as expected from Eq.(1). Using Eq. (1), we can approximate $\Gamma_{I,short}$ as, 
\begin{eqnarray}
  \Gamma_{I,short} \approx (1+\kappa_I)\frac{1}{A}\int_A C_{I^\prime,short}(\Delta r) d^2(\Delta r),
\end{eqnarray}
in which, $\frac{1}{A}\int_A C_{I^\prime,short}(\Delta r) d^2(\Delta r)$ corresponds to purely Gaussian random fluctuation.

Equation (3) cannot be applied directly to $v$ because the singularities do not exist at every point as does the intensity. Finding singularities separated by $\Delta r$ must be described as a correlated random process with a probability which is not uniform in $\Delta r$. Thus ${\rm var}(\tilde{v}_a)$ cannot be expressed simply as a two-dimensional integral of $C_v(\Delta r)$ as was the case for intensity. However, the short-range contribution to ${\rm var}(\tilde{v}_a)$ can be evaluated using the measured values of ${\rm var}(\tilde{v}^\prime_a)$, which corresponds to fluctuations for Gaussian waves, multiplied by the mesoscopic enhancement factor $(1+\kappa_v)$ as in Eq. (4) for the intensity. We then expect that 
\begin{eqnarray}
  {\rm var}(\tilde{v}_a) \approx (1+\kappa_v){\rm var}(\tilde{v}^\prime_a) +\kappa_v.
\end{eqnarray}
In the limit $\kappa_v\rightarrow 0$, ${\rm var}(\tilde{v}_a)$ reduces to the Gaussian term ${\rm var}(\tilde{v}^\prime_a)$. Using values of ${\rm var}(\tilde{v}^\prime_a)$ in Table 1, Eq. (5) gives ${\rm var}(\tilde{v}_a)\approx 0.175$ and 1.09, for diffusive and localized waves, which are in reasonable agreement with the corresponding measured values, 0.193 and 1.24. 

\begin{figure}
\includegraphics[scale=0.7]{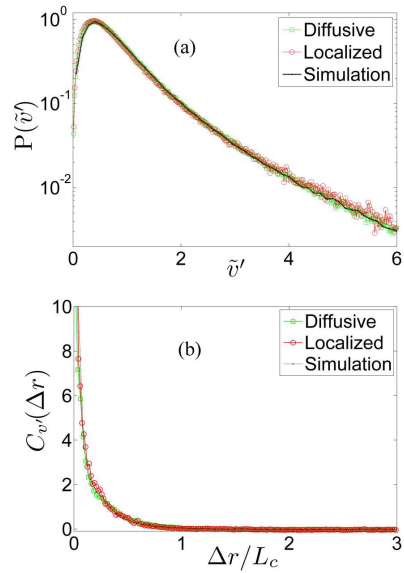}
\caption{\label{Fig3} (Color Online) Probability distributions (a) and the cumulant spatial correlation functions (b) of the normalized velocity of phase singularities $v^\prime = v/\sigma_{\Delta I^*}$ for diffusive and localized waves and the comparison to simulations for Gaussian random waves.}
\end{figure}

In conclusion, we have demonstrated a unified framework for the statistics of transmission and speckle change. In each case, mesoscopic fluctuations disappear when the local variable is normalized by a global variable reflecting the speckle pattern as a whole. In the limit in which a large number of modes contribute to the field, first and second order statistics of the normalized local variable approach the statistics of a Gaussian random process. We expect that the statistics of change in the speckle pattern, which may arise from internal motion of the sample, temperature change, time delay following pulsed excitation or by non-monochromatic excitation can be described within this framework.

We thank Bing Hu for contributions to the experiment and data analysis. This work was supported by the NSF under Grant No. DMR-0538350.

\end{document}